\documentclass[prl,twocolumn,superscriptaddress,showpacs,a4paper]{revtex4}%
\usepackage{epsfig}
\usepackage{amsmath}
\usepackage{graphicx}
\usepackage{bm}
\usepackage{amsfonts}
\usepackage{amssymb}%
\begin{document}
\title{Room-temperature steady-state optomechanics entanglement on a chip}
\author{Chang-Ling Zou}
\affiliation{Key Laboratory of Quantum Information, University of Science and Technology of
China, Hefei, Anhui 230026, P. R. China.}
\author{Xu-Bo Zou}
\email{xbz@ustc.edu.cn}
\affiliation{Key Laboratory of Quantum Information, University of Science and Technology of
China, Hefei, Anhui 230026, P. R. China.}
\author{Fang-Wen Sun}
\affiliation{Key Laboratory of Quantum Information, University of Science and Technology of
China, Hefei, Anhui 230026, P. R. China.}
\author{Zheng-Fu Han}
\email{zfhan@ustc.edu.cn}
\affiliation{Key Laboratory of Quantum Information, University of Science and Technology of
China, Hefei, Anhui 230026, P. R. China.}
\author{Guang-Can Guo}
\affiliation{Key Laboratory of Quantum Information, University of Science and Technology of
China, Hefei, Anhui 230026, P. R. China.}
\date{\today}

\begin{abstract}
A potential experimental system, based on the Silicon Nitride (SiN)
material, is proposed to generate steady-state room-temperature
optomechanical entanglement. In the proposed structure, the
nanostring interacts dispersively and reactively with the microdisk
cavity via the evanescent field. We study the role of both
dispersive and reactive coupling in generating optomechanical
entanglement, and show that the room-temperature entanglement can be
effectively obtained through the dispersive couplings within the
reasonable experimental parameters. In particular, we find, in the
high Temperature ($T$) and high mechanical qualify factor ($Q_{m}$)
limit, the logarithmic entanglement depends only on the ratio
$T/Q_{m}$. This means that improvements in the material quantity and
structure design may lead to more efficient generation of stationary
high-temperature entanglement.

\end{abstract}

\maketitle

\emph{Introduction.-} The quantum entanglement not only represents one of the
most interesting fundamental phenomena in quantum mechanics, but also is
regarded as very important resource in quantum computation and quantum
information processing \cite{qi}. It has been a focus in quantum mechanics
since the seminal work of Einstein-Podolsky-Rosen (EPR) gedanken experiment
\cite{epr}. Nowadays, quantum entanglement has been observed in various
quantum systems, such as photons, atoms, and quantum dots \cite{entanglements}%
. Although it is believed that large decoherence effect would mask quantum
signatures of macroscopic objects, great efforts are still dedicated to
macroscopic quantum entanglement, in order to exploring the boundary between
quantum and classical mechanics.

Recently, rapid progress in nanofabration technologies offers novel
opportunities to study macroscopic quantum entanglement \cite{Schwab}.
Especially, combined with mature optical technologies, the emerging
optomechanics enable precisely controlling the motion of mechanical
oscillators through optical radiation force or gradient force \cite{om}. It
was proposed that the mechanical vibration could be entangled with the optical
field in various optomechanics systems, such as Fabry-P\'{e}rot cavity with a
movable mirror \cite{Vitali1} or nanomembranes \cite{Hartmann}. It could also
be extended to microwave cavities \cite{Vitali2}. Unfortunately, quantum
entanglement in these systems is very sensitive to the temperature, typically
only valid in cryostat. Up to now quantum entanglement related to mechanical
modes has not been observed in any experiment. One of the key question arises:
Is it possible to entangle a light beam and macroscopic objects at room temperature?

\begin{figure}[ptb]
\centerline{
\includegraphics[width=0.45\textwidth]{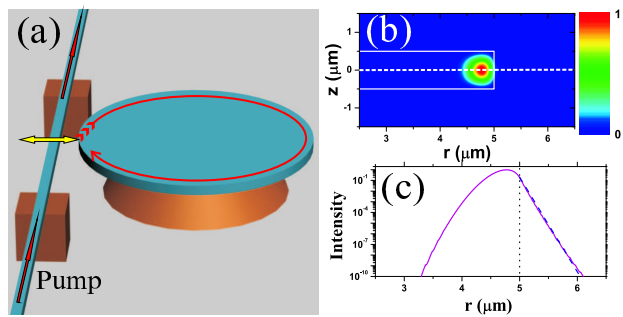}}\caption{(color online) (a)
Schematic illustration of a nanostring coupled to a high-Q WG
microcavity. The microcavity is evanescent-field driven by the
waveguide, which is also act as a high-Q mechanical oscillator. (b)
The normalized field $\left\vert E\right\vert ^{2}$ distribution of
cross section of the SiN microdisk with diameter $\Phi=10\mu m$ and
thickness $t=1\mu m$ (c) The normalized field $\left\vert
E\right\vert ^{2}$ distribution at $z=0$ (dashed line in (b)), the
evanscent field is
fit to a exponential decay curve with decay length $l_{0}=100nm$.}%
\end{figure}

In this paper, we propose a potential microcavity-nanostring system
to generate the room-temperature optomechanical entanglement. The
system is based on the Silicon Nitride material, the nanostring
interacts dispersively and reactively with the microdisk cavity via
the evanescent field. It is shown that  entanglement can be
effectively generated through dispersive couplings within the
current experimental parameters, and can be preserved at room
temperature or even at a high temperature. We analyzes the the
dependence of entanglement on the temperature and mechanical
dissipation, and find that the logarithmic entanglement is function
of the ratio $T/Q_{m}$ in the high-$Q_{m}$ and high-temperature
limit $T$. This demonstrates that the improvements in the material
quantity and structure design may lead to more efficient generation
of stationary room-temperature entanglement.

\emph{System.-} As shown in Fig.1(a), the system consists of a
nanostring oscillator and a microdisk cavity, fabricated by SiN,
which can be integrated on single chip. This structure have been
experimentally realized in silicon chip, but the performance is
limited by low mechanical quality factor ($Q_{m}$) and low optical
quality factors ($Q_{o}$) \cite{Li}. Similar structure with a
nanostring closed to a silica toroid has also been studied by
Anetsberger et al.\cite{Anet}, which shows displacement sensitivity
beyond the standard quantum limit. However, it is a challenge to
precisely control the gap between the nanostring and cavity in
experiments\cite{Anet}. Compared with these studies\cite{Li,Anet},
the experimental system proposed in this paper has the following
advantages: (1) The fabrication technology of SiN device is
compatible with silicon, and permits further expendability of
on-chip optical components and opto-electronics elements. (2) The
SiN is transparent for a wide band, and has relative high refractive
index. Thus dielectric microcavity based on SiN can realize the
great confinement of light, possess ultrahigh  qualify factor and
small mode volumes. Currently, the whispering gallery (WG) modes
with qualify factor $Q_{o}$ up to about $4\times10^{6}$ have been
demonstrated experimentally\cite{SiNopt}. (3) The mechanical
properties of SiN is excellent, so that we can fabricate strained
nanostrings with mechanical qualify factor $Q_m>10^{6}$ at room
temperature\cite{SiNstring}. The large $Q_{o}$\ significantly
enhanced the light-matter interaction at low input intensity, and
large $Q_{m}$ permits very long coherence time. Therefore, our
system provides a unique combination of ultrahigh mechanical and
optical quality factor, and holds great potential for quantum
optomechanics.

To study the interaction between the microcavity and nanostring, we
plot the mode field distribution of microdisk in Fig.1(b), which
with diameter $\Phi=10\mu m$ and thickness $t=1\mu m$. It is shown
that the WG modes is well confined in the microdisk, with radiation
related $Q>10^8$. A small portion of energy outside the cavity form
the evanescent waves, which is exponentially decay with the distance
from the dielectric interface $(x)$, and can be approximately
describe by $E_{c}(x)=E_{c} (0)e^{-x/l_{0}}$, where $l_{0}$ is the
decay length. In Fig.1(c), we plot the normalized field $\left\vert
E\right\vert ^{2}$ distribution at $z=0$, and find that the
evanscent field is fit to a exponential decay curve with decay
length $l_{0}=100nm$. In addition, as shown in Fig.1(a), the
nanostring, as a waveguide, transfers light into and out from the
microcavity. When the nanostring is placed in the evanescent-field
of the WG mode, the mechanical oscillator will be attracted to the
cavity through the gradient force, and the displacement of the
nanostring also gives a backaction on the WG mode, and changes both
the resonance frequency (dispersive coupling, DC) and energy decay
rate (reactive coupling, RC).

From the coupling mode theory \cite{Gorodetsky}, the frequency shift
$\delta\omega$ to the cavity resonance and the coupling strength $\kappa_{1}$
between the cavity mode and waveguiding mode can be expressed as $\delta
\omega(x)\approx\delta\omega(0)e^{-2x/l_{0}}$, and $\kappa_{1}(x)\approx
\kappa_{1}(0)e^{-2x/l_{0}}$. Thus, for a small displacement $x$ around steady
position, $\omega_{c}(x)\approx\omega_{c}(0)-dx,$ and $\kappa_{1}%
(x)\approx\kappa_{1}(0)+rx,$where $d$ and $r$ describe the dispersive and
reactive coupling strengths between the mechanical and optical modes,
respectively, and both of them could be controlled experimentally by adjusting
the size of disk and nanostring and gap between them.

\emph{Model.-} The Hamiltonian of the\ coupled microcavity and nanostring
system is given by \cite{Qoptics,Gio}
\begin{align}
H  &  =\hbar\omega_{c}a^{\dag}a+\frac{1}{2}\hbar\omega_{m}(p^{2}+q^{2})-\hbar
Da^{\dag}aq\\
&  +i\hbar(\sqrt{2\kappa_{1}(x)}+Rq/\sqrt{2\kappa_{1}(x)})E(e^{-i\omega_{l}%
t}a^{\dag}-e^{i\omega_{l}t}a),\nonumber
\end{align}
where Bose operators $a$ and $a^{\dag}$ represent the annihilation and
creation operators of the cavity mode with frequency $\omega_{c}$ and
intrinsic loss $\kappa_{0}$. For the nanostring, we define dimensionless
position and momentum operators $q=\sqrt{m\omega_{m}/\hbar}x$, $p=\sqrt
{m/\hbar\omega_{m}}\dot{x}$, where $m\ $stands for effective mass of the
nanostring with a resonant frequency $\omega_{m}$. The corresponding
dispersive and reactive coupling strength are normalized by zero point
fluctuation $\sqrt{\hbar/m\omega_{m}}$: $D=d/\sqrt{m\omega_{m}/\hbar}$ and
$R=r/\sqrt{m\omega_{m}/\hbar}$. The system is driven by a coherent laser with
frequency $\omega_{l}$ and power $P$, corresponding to the driving field $E$
$=\sqrt{P/\hbar\omega_{l}}$.

By including the noise on both the mechanical ($\xi(t)$) and optical
($a_{in}(t)$) modes, we obtain the quantum Langevin equations (QLE) in the
reference frame rotating with frequency $\omega_{l}$,%
\begin{subequations}
\begin{align}
\dot{q}= &  \omega_{m}p,\\
\dot{p}= &  -\omega_{m}q-\gamma_{m}p+Ga^{\dag}a-iR/\sqrt{2\kappa_{1}}%
E(a^{\dag}-a)\nonumber\\
&  -iR/\sqrt{2\kappa}(a^{\dag}a_{in}-aa_{in}^{\dag})+\xi(t),\\
\dot{a}= &  -i\Delta a-(\kappa+Rq)a+iGqa\nonumber\\
&  +(\sqrt{2\kappa_{1}}+Rq/\sqrt{2\kappa_{1}})E+\sqrt{2\kappa}a_{in}(t),
\end{align}
where $\kappa=\kappa_{0}+\kappa_{1}$,
$\Delta=\omega_{c}-\omega_{l}$, and the noise correlation function
are $\left\langle a_{in}(t)a_{in}^{\dag}(t^{\prime })\right\rangle
=\delta(t-t^{\prime})$, and $\left\langle \xi(t)\xi(t^{\prime
})\right\rangle =\gamma_{m}(2\bar{n}+1)\delta\left(
t-t^{\prime}\right)  $. Here $\bar{n}=1/\left(  \exp\left(
\hbar\omega_{m}/k_{B}T\right)  -1\right) $ is the mean thermal
phonon number, where $k_{B}$ and $T$ denote Boltzmann constant and
temperature. By linearizing operators around the steady state values
($o=o_{s}+\delta o$, $o$ is a system operator), we obtain the QLE
for the fluctuation operators
\end{subequations}
\begin{equation}
\dot{f}(t)=Mf(t)+n(t),
\end{equation}
where $f=(q,p,X,Y)^{T}$ is the vector of fluctuation operators$,n$ is
corresponding noises and the matrix
\begin{equation}
M=\left(
\begin{array}
[c]{cccc}%
0 & \omega_{m} & 0 & 0\\
-\omega_{m} & -\gamma_{m} & DX_{s} & -R\frac{E}{\sqrt{\kappa_{1}}}\\
R\frac{E}{\sqrt{\kappa_{1}}}-RX_{s} & 0 & -\kappa_{s} & \Delta_{s}\\
DX_{s} & 0 & -\Delta_{s} & -\kappa_{s}%
\end{array}
\right)  .
\end{equation}
Here we introduce the cavity field quadratures $X=(a+a^{\dag})/\sqrt{2}$ and
$Y=i(a^{\dag}-a)/\sqrt{2}$. By carefully choosing the phase of driving light,
we set Im$a_{s}=0$, then $X_{s}=\sqrt{2}a_{s}$ and $Y_{s}=0$. And $\Delta
_{s}=\Delta-Dq_{s}$, $\kappa_{s}=\kappa+Rq_{s}$ the detuning and decay rate of
cavity mode at steady state.

When stability condition is fulfilled, which can be derived by
employing the Routh-Hurwitz criterion \cite{Routh}, we can solve the
stochastic differential
equation for the steady-state correlation matrix ($V$)%

\begin{equation}
M\cdot V+V\cdot M^{T}=-I,
\end{equation}
where $V$ is defined as $V_{ij}=\left\langle f_{i}f_{j}+f_{j}f_{i}%
\right\rangle /2$, and $I=Diag\left[
0,\gamma_{m}(2\bar{n}+1)+(R/\sqrt
{2\kappa_{1}})^{2}X_{s}/2,\kappa,\kappa\right]  $. We represent $V$
in the $2\times2$ block form,

\begin{equation}
V=\left(
\begin{array}
[c]{cc}%
A & C\\
C^{T} & B
\end{array}
\right)  ,
\end{equation}
then the entanglement of between optical field and mechanics oscillation can
be quantified by the logarithmic negativity $E_{N}$ \cite{EN}:
\begin{equation}
E_{N}=Max[0,-\frac{1}{2}\ln2(\Sigma-\sqrt{\Sigma^{2}-4\det V})],
\end{equation}
where $\Sigma=\det A+\det B-2\det C$.

\begin{figure}[ptb]
\centerline{
\includegraphics[width=0.45\textwidth]{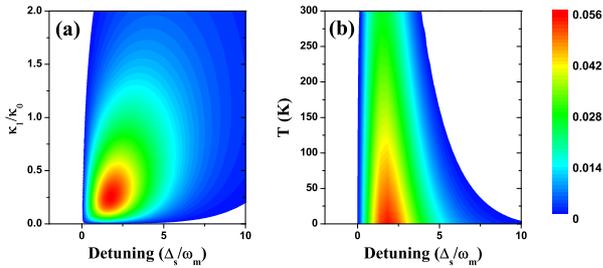}}\caption{(color online)
Logarithmic negativity $E_{N}$ against coupling parameter $\kappa_{1}%
/\kappa_{0}$, temperature $T$ and detuning $\Delta_{s}$, with $T=0.05K$ (a),
$\kappa_{1}/\kappa_{0}=0.3$ (b). The blank region means $E_{N}=0$, the input
power is fixed at 100mW.}%
\end{figure}

\emph{Results and Discussion.-} To model the quantum entanglement
between light field and motion of nanostring, we adapt the
parameters closed to recent experiments. The microdisk with
$Q_{o}=4\times10^{6}$ is working at wavelength $\lambda=850$ nm,
which have been demonstrated experimentally\cite{SiNopt}. The
nanostring is chosen with $\omega_{m}=15$ MHz, $m=2$ pg, and
$Q_{m}=10^{6}$ at room temperature, which is achievable by current
technology \cite{SiNstring}. We set the coupling strength
$r=2\kappa_{1}/l_{0}$ with $l_{0}=100$\textrm{ nm}, and $g=50$
MHz$/$nm. All calculations of $E_{N}$ are performed under the
stability condition.

In Fig. 2(a), we calculated the dependence of $E_{N}$ on the cavity
mode detuning $\Delta_{s}$ and the coupling condition
$\kappa_{1}/\kappa_{0}$ at low temperature ($T=0.05$ \textrm{K}). It
can be seen that entanglement appears at blue detuning, and there
exist a optimal coupling strength $\kappa_{1}/\kappa_{0}\approx0.3$
for entanglement. This can be understood as follows: From the
Eq.(4), we see that the entanglement depends on effective
opto-mechanics coupling
strength $\zeta=D\frac{\sqrt{2\kappa_{1}}E}{i\Delta_{s}+\kappa_{1}+\kappa_{0}%
}$ and $\eta=R\frac{E}{\sqrt{\kappa_{1}}}$. If the stationary
condition is fulfilled, we could say, the entanglement is enhanced
by increasing $\zeta$ and $\eta$. Thus, there is a optimal coupling
strength as a result of the trad-off between the energy dissipation
and transferring introduced by the waveguide.

\begin{figure}[ptb]
\centerline{
\includegraphics[width=0.45\textwidth]{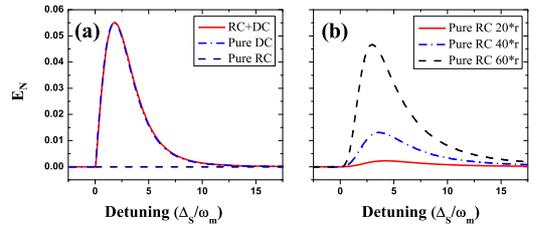}}\caption{(color online)
$E_{N}$ vs. detuning $\Delta_{s}$ for (a) realism condition contain both
reactive coupling (RC) and dispersive coupling (DC), compares to the pure RC
with $g=0$ and DC with $r=0,$ (b) pure RC with $g=0$, enhanced reactive
coupling strength by 20, 40 and 60 times. The input power $P=100$ mW,
temperature $T=0.05$ K and $\kappa_{1}/\kappa_{0}=0.3$.}%
\end{figure}

Since there are two different kinds of interactions (DC and RC) in
our system, we should investigate the role of them in generating
entanglement. As seen from Fig.3(a), we compares the $E_{N}$ in
three different conditions: (1) both DC and RC are existing, (2)
only DC and (3) only RC. The results demonstrated that, the RC do
not influence or generate the entanglement in the
microdisk-nanostring system proposed in this paper. However, if we
can increase the RC coupling strength $r$, as shown in Fig.3(b), the
entanglement appears. This means that the RC could also lead to
entanglement, which would be significant in other systems which have
strong RC coupling.

In Fig. 2(b), we study the temperature dependence of $E_{N}$. Here,
we choose $Q_{m}=10^{6}$ at room temperature, which have been
demonstrated in experiments \cite{SiNstring}. It can be seen that at
the optimal coupling condition, with increasing the temperature
($T$), the $E_{N}$ decreases and the working area is also shrank. It
is surprised that, the entanglement is robust against the
temperature, even at room temperature (300 \textrm{K}), the
optomechanical entanglement can be well preserved. In other words,
the macroscopic oscillator can be entangled with a cavity mode field
at room temperature by carefully choosing the coupling condition and
frequency of pumping light. It seems quite wired that the
entanglement could be preserved at such high temperature, because
when the enviroment temperature largely exceeds the energy of the
one quantum, i.e. $k_{B}T>\hbar\omega_{m}$, the quantum signature
would be masked by the thermal noise.  This counterintuitive
phenomena can be understood as follow: The system's interaction with
the thermal bath is governed by the mechanical dissipation rate
$\gamma_{m}=\omega_{m}/Q_{m}$. When the interaction between the
light and oscillator is much bigger than the rate of relaxation to
thermal equilibrium, the quantum entanglement could be survived. If
we assume $\gamma_{m}=0$ $(Q_{m}=\infty)$,  the quantum system is
total isolated from the thermal bath, and the quantum behavior could
be preserved at any temperature.

To study the dependence of entanglement on $Q_{m}$, in Fig.4(a), we
further plot $E_{N}$ against $T$ with different $Q_{m}$, by assuming
all parameters do not change with temperature. Benefitting from
large $Q_{m}=10^{6}$ of SiN nanostring in our system, the
entanglement vanished at a very high critical temperature $T_{C}$.
However, for lower $Q_{m}$, $T_{C}$ is decreased, which corresponds
to former results that entanglement only survived in cryostat
\cite{Vitali1}. It is not difficult to find in Fig.4(a) that, all
curves are the same shape for different $Q_{m}$.

In order to get a better physical picture, we analysis Eq.(4) and
Eq.(5) qualitatively. When $Q_{m}$ is very large, $\gamma_{m}/\omega
_{m}=1/Q_{m}\ll1$, and $\gamma_{m}/\kappa\ll1$, we can approximately
omit the $\gamma_{m}$ in matrix $M$ of Eq.(4). In this simplified
model, from equation (5), we can obtain the dependence of solution
of correlation matrix ($V$) on the temperature $T$, which is only
related to $I$, and depends on the parameter
$U(Q_{m},T)=\gamma_{m}(2\bar{n}+1)\approx2\frac{1}{\left(
\exp\left(  \hbar\omega_{m}/k_{B}T\right)  -1\right)  }\frac{\omega_{m}}%
{Q_{m}}$. At low temperature $T\approx0$, $U(Q_{m},0)\approx0$, we can see
from Fig.4(a) that $E_{N}$ is not related to the $Q_{m}$. At high temperature,
i.e. $T\gg1$, $U(Q_{m},T)\approx\frac{k_{B}}{\hbar}\frac{T}{Q_{m}}$, then
$E_{N}$ is a function with $\frac{T}{Q_{m}}$. In Fig. 4(b), the dependence of
$T_{C}$ of entanglement on the $Q_{m}$ shows great agreement with the
relationship $\frac{T}{Q_{m}}=const$ when $Q_{m}>100$.

\begin{figure}[ptb]
\centerline{
\includegraphics[width=0.45\textwidth]{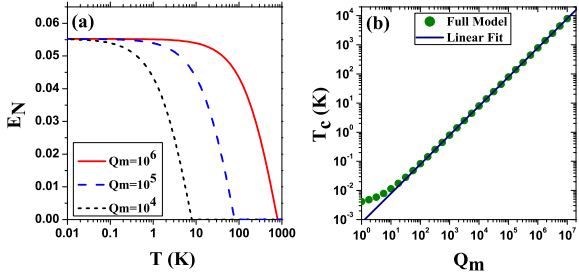}}\caption{(color online) (a)
$E_{N}$ vs the temperature $T$, with different $Q_{m}$. (b)The critical
temperature $T_{C}$ as a function of $Q_{m}$, the line is a linear fit to the
full model data. Parameters are $P=100$ mW, $\kappa_{1}/\kappa_{0}\approx0.3$
and $\Delta_{s}=1.8\omega_{m}$.}%
\end{figure}

\emph{Conclusion.-} A potential experimental system has been
proposed to generate stationary room-temperature optomechanical
entanglement. The system consists of a high-Q nanostring oscillator
and a microdisk cavity, fabricated by SiN, which can be integrated
on single chip. The role of both dispersive and reactive coupling in
generating optomechanical entanglement is studied. We find that in
the system proposed in this paper, the room-temperature entanglement
is effectively obtained through the dispersive couplings. In
particular, we find, in the high Temperature ($T$) and high
mechanical qualify factor ($Q_{m}$) limit, the logarithmic
entanglement depends only on the ratio $T/Q_{m}$. This means that
improvements in material quality, and optimization in the structure
design may enable both $Q_{o}$ and $Q_{m}$ greater, and lead to more
efficient generation of stationary entanglement. This system can be
applied as fundamental elements on photon chips\cite{Tian,
Stannigel,OBrien}. With a rapid development in material science and
fabrication technology, the scheme proposed here is expected to be
realized in the near future, and to explore the high-temperature
entanglement of macroscopic objects \cite{add}.

\emph{Acknowledgement.-} We gratefully acknowledge Ming Gong, Yong Yang,
Zhang-Qi Yin and Chun-Hua Dong for fruitful discussion, and Prof. Yun-Feng
Xiao at Peking University for insightful comments on our manuscript. This work
is supported by the National Natural Science Foundation of China under Grant
No. 11074244 and No. 11004184.

\end{document}